# Optical Companding

**Yunshan Jiang, Bahram Jalali**
*Electrical and Computer Engineering, UCLA, Los Angeles, CA*
yunshanjiang@gmail.com

**Abstract:** We introduce a new nonlinear analog optical computing concept that compresses the signal's dynamic range and realizes non-uniform quantization that reshapes and improves the signal-to-noise ratio in the digital domain.

## 1. Optical non-uniform quantization with optical companding

In nearly all optical sensing applications, the trade-off between speed, sensitivity and full-scale poses fundamental challenges to the detection [1-3]. Optical amplifiers boost the lower portion of dynamic range over the noise floor help the measurement of small amplitudes, but it also increases the full-scale and requires the receivers to have a larger dynamic range and higher number of bits for quantization [4].

Analog optical computing promises to alleviate bottlenecks associated with digital acquisition and processing of optical data [5,6]. Optical companding proposed here tackles the trade-off between dynamic range and sensitivity by boosting the small amplitudes over the noise floor, while keeping the full-scale the same. This relieve the requirement on quantizing resolution. The concept is illustrated in Fig. 1. The dynamic range of the input optical signal is compressed in the optical dynamic range compressor, whose gain changes dynamically as a logarithmic-like function of the instantaneous input power. After detected by the photodiode and digitized by the analog-to-digital-converter (ADC), the compressed digital signal is recovered through the digital expander that is the inverse mapping function of the logarithmic gain profile of optical compression. Without requiring an optical receiver with larger dynamic range or more quantization bits, the detectable range is extended.

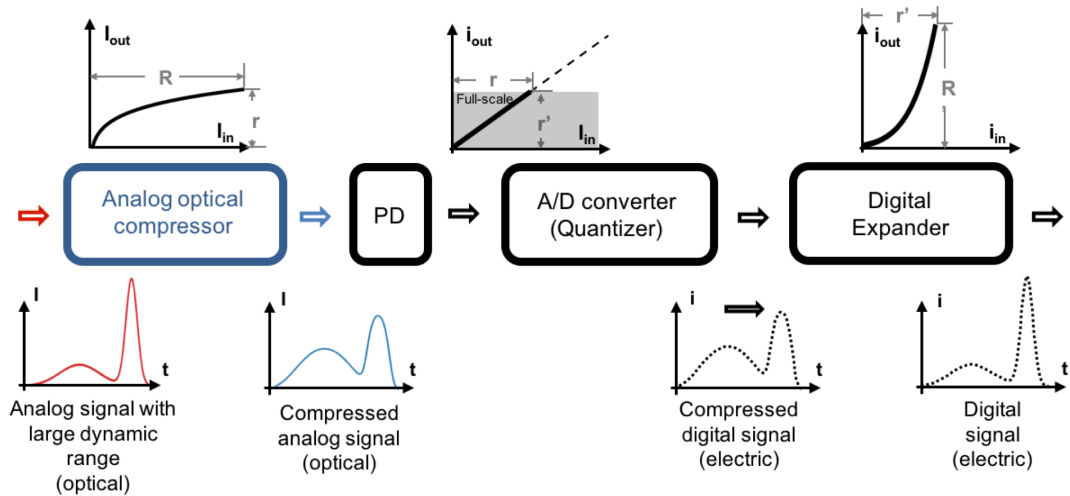

Fig. 1. Diagram of the Optical Companding. PD, photodiode.

The combination effect of the optical compressor and the linear quantizer results in the non-uniform quantization where quantization step increases as a function of the input signal amplitude. In the quantization process, the amplitude of the analog input is classified into non-overlapping bins. Each bin is then mapped to a reconstruction value proportional to the input. Fig. 2 demonstrates the output of a uniform ADC when the signal is reshaped in optical companding. Compared to the linear case, the lower portion of the dynamic range is quantized with higher resolution, at the expenses of the accuracy of the upper portion. This allows higher signal-to-noise ratio for small inputs, while trading-off the excess SNR for the strong peaks. In the applications where signals with strong peaks are relatively rare, less bits are required to quantize the optically-companded signal for the similar SNR.

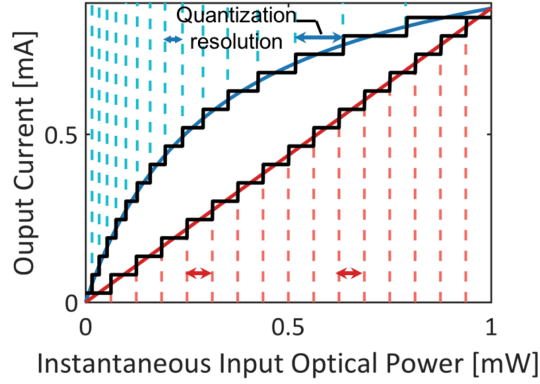

Fig. 2 Optical companding enables non-uniform quantization. Compared to the linear case (red curves), the lower part of the dynamic range is associated with finer quantization resolution and higher signal-to-noise ratio (for low inputs). The logarithmic-type compression (blue curve) is obtained using nonlinear propagation in silicon with nonlinear losses and saturated Raman amplification. Similar behavior can be achieved with other types of amplifiers (with fast gain response) as well as with nonlinear refraction (see Fig. 4).

The mean square quantization error, or quantization noise power of an ideal ADC is calculated as [7,8]:

$$\sigma_q^2 = E\left[(x - Q(x))^2\right] = \int_{-\infty}^{\infty}(x - Q(x))^2 p(x)dx$$

$$= \sum_{k=1}^{2^n}\int_{b_{k-1}}^{b_k}(x - y_k)^2 p(x)dx \qquad \text{eq. (1)}$$

where $x$ is the input signal, $Q: x \to y_k, \forall\, x \in [b_{k-1}, b_k)$ is the mapping function of quantization, $p(x)$ is the probability density function of $x$, and n is the bit number of the ADC. When the quantization step $\Delta_k = b_k - b_{k-1}$ is small enough so that $p(x)$ is approximately uniform within each bin, eq. (1) can be approximated as:

$$\sigma_q^2 \approx \sum_{k=1}^{2^n}\frac{\Delta_k^3}{12}\int_{b_{k-1}}^{b_k}p(x)dx = \frac{1}{12}\sum_{k=1}^{2^n}\Delta_k^3 \cdot p(x \in [b_{k-1}, b_k)) \qquad \text{eq. (2)}$$

where $p(x \in [b_{k-1}, b_k))$ is the probability of the input falling in a given quantization bin.

In uniform quantizer, the quantization resolution $\Delta$ is a function of the maximum range of ADC and the ADC bit number n, and is thus signal-independent,

$$\Delta = \frac{V_{FS}}{2^n} \qquad \text{eq. (2)}$$

The uniform quantizer minimize the quantization noise power only when the probability of the input falling in each interval is the same, resulting in

$$\sigma_q^2 = \frac{\Delta^2}{12} \qquad \text{eq. (3)}$$

Unlike the linear case, optical companding assign more bits for the more frequent small amplitudes. The quantization resolution as a function of input power is calculated as:

$$\Delta_{OC}(P_{in}) = \frac{\Delta_{linear}}{g'(P_{in})} = \frac{1}{g'(P_{in})} \cdot \frac{V_{FS}}{2^n} \qquad \text{eq. (4)}$$

Where $g'()$ is the output amplitude as a function of the input power.

The quantization resolution as a function of input power is calculated as:

$$\sigma_q^2 \approx \frac{1}{12}\sum_{k=1}^{2^n}\left(\frac{\Delta}{g'(P_{in})}\right)^3 p(P_{in}) \qquad \text{eq. (5)}$$

In the applications where signals with strong peaks are relatively rare, the optical companding reshapes the quantization noise and improve the SNR. The reshaping of quantization noise is shown in Fig. 3.

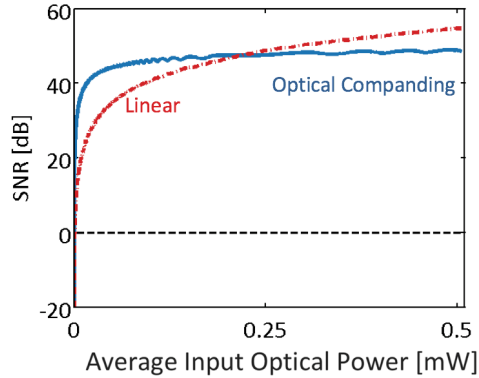

Fig. 3 The optical companding reshapes quantization noise. As the combined effect of logarithm shaping and linear quantization, small input is associated with small quantization noise and higher SNR, at the expense of the excess SNR at large input.

## 2. Implementation of optical companding

Optical companding devices must be fast to response to the input's instantaneous temporal envelop and have a non-uniform gain/loss that decreases/increases as the input amplitude increases. The diagram of companding mechanisms based on saturated amplification, nonlinear absorption, and nonlinear refraction is shown Fig. 3.

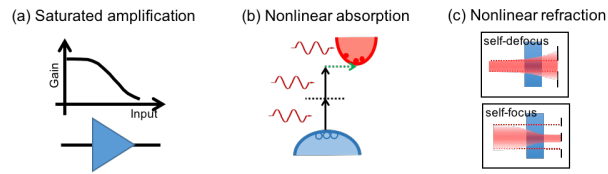

Fig. 4 Diagrams of optical companding mechanisms.

Saturated amplifiers can be used to shape the input amplitude by providing a higher gain for the small inputs, and lower gain for the large inputs as the pump is depleted. This logarithmic-type transfer function can be tuned by changing the pump power. A stronger pump leads to a smaller saturation power. Since it is an inherently accumulative process, the gain relaxation of the medium needs to be faster than the dynamics of the signal. Nonlinear optical amplification and saturated semiconductor optical amplifiers are candidates for implementing the proposed companding. Saturated optical Raman amplification in silicon photonics [9] has been demonstrated for optical computation based on logarithmic operations [10].

Two-photon absorption (TPA) provides loss that increases quadratically with the input intensity and can be used to compress the dynamic range. The process has the desired ultrafast response. Submicron-optical waveguides confine the input power in submicron area and greatly enhance the absorption. In silicon photonics, absorption from the TPA-generated free carriers (FCA) also contributes to the attenuation of the large input, as shown in Fig. 3(b) (modified from ref. [11]).

The logarithm-like transform function in Fig.2 is modeled by the nonlinear propagation in a silicon photonic waveguide with the consideration of the saturated Raman amplification, TPA and FCA. The input pump is 7mW, the waveguide length is 2cm and the mode area is $0.1um^2$. The phenomenon is elaborated in Ref. [10].

Nonlinear refraction mechanisms form another class of optical companding devices. When the high-power input enters the nonlinear material, the change in refractive index causes the beam to self-focus or self-defocus and be blocked by a system aperture places at the output. As opposed to the absorption-based devices, such methods have a larger damage threshold and can be used for optical limiting [12].

## 3. Conclusion

Optical companding is a new analog optical computing method that logarithmically shapes the input signal amplitude and relieves the requirement on the dynamic range of the detector and the A/D converter. It reshapes the quantization noise and improves the SNR in the digital capture of the weak inputs. While such nonlinear transformation increases the signal bandwidth, they will improve performance when the system is dynamic range limited. On the other hand, time stretch techniques [13] can be employed if the system becomes bandwidth limited.

This work was supported by the Office of Naval Research (ONR) MURI Program on Near-Field Nanophotonics for Energy Efficient Computing and Communication (NECom).